# WAVEFORM FEATURE IMPLEMENTATION FOR FRIB LLRF CONTROLLERS*

S. Zhao†, E. Bernal-Ruiz, M. Konrad, Z. Li, H. Maniar, D. Morris, Facility for Rare Isotope Beams, Michigan State University, East Lansing, MI, USA

*Abstract*

Waveform feature is one of the requirements for the FRIB LLRF controllers. It is desired that the LLRF controllers store the internal data (e.g. the amplitude and phase information of forward/reverse/cavity signals) for at least one second of sampled data at the RF feedback control loop rate (around 1.25 MHz). One use case is to freeze the data buffer when an interlock event happens and read out the fast data to diagnose the problem. Another use case is to monitor a set of signals at a decimated rate (user settable) while the data buffer is still running, like using an oscilloscope. The detailed implementation will be discussed in the paper, including writing data into the DDR memory through the native interface, reading out the data through the bus interface, etc.

## INTRODUCTION

The Facility for Rare Isotope Beams (FRIB) is a scientific user facility for nuclear physics research being built at Michigan State University (MSU). The low level radio frequency (LLRF) controller for the FRIB project is designed to accommodate various cavity and tuner types [1] (see Table 1).

Table 1: FRIB Cavity Types

| System | Area | Frequency (MHz) | Type | Tuner |
|---|---|---|---|---|
| MHB F1 | FE | 40.25 | RT | N/A |
| MHB F2 | FE | 80.5 | RT | N/A |
| MHB F3 | FE | 120.75 | RT | N/A |
| RFQ | FE | 80.5 | RT | temperature |
| MEBT | FE | 80.5 | RT | 2-phase stepper |
| QWR | LS1<br>FS1 | 80.5 | SC | 2-phase stepper |
| MGB | FS1 | 161 | RT | 5-phase stepper |
| HWR | LS2/3<br>FS2 | 322 | SC | pneumatic |

With the direct-sampling and under-sampling techniques the analog-to-digital converter (ADC) can sample RF signals of five different frequencies (in harmonic relation) ranging from 40.25 MHz to 322 MHz with the same sampling frequency. Non-In-Phase/Quadrature (non-IQ) sampling is also adopted to reduce the effect of harmonics. RF output synthesis is done by generating four points (I, Q, −I, −Q) per waveform and using band-pass filters to pick up the fundamental or higher harmonics of interest. There are three RF board variations, and each supports two frequencies. Two types of tuner boards of the same form factor are designed to drive stepper motors or analog tuners.

* Work supported by the U.S. Department of Energy Office of Science under Cooperative Agreement DE-SC0000661
† Email address: zhaos@frib.msu.edu

The FRIB LLRF controller includes: the FRIB general purpose digital board (FGPDB), one of the three RF board variations, one of the two types of tuner boards or none, a front panel board, a power supply and a chassis (see Fig. 1).

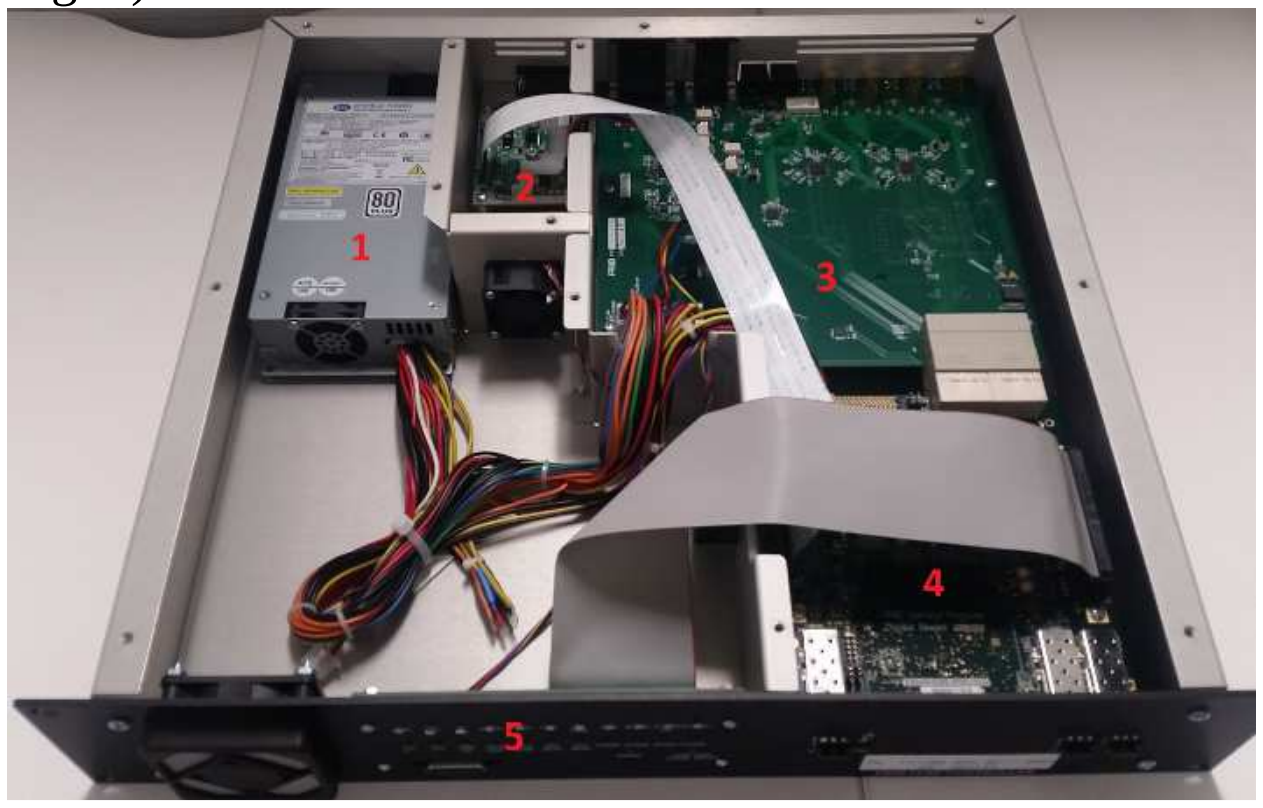

Figure 1: FRIB LLRF controller. 1. Power supply; 2. Tuner board; 3. RF board; 4. FGPDB; 5. Front panel and board.

Besides flexibility, the FRIB LLRF controller also features: 1. digital self-excited loop (SEL) allowing cavities to be driven with tuners unlocked; 2. advanced control algorithms for high performance [2]; 3. comprehensive external and internal interlocks to protect the cavities, amplifiers, and other equipment; 4. auto-start procedure to automate cavity turn-on for efficient operation; 5. waveform storage for monitoring and diagnosis.

## FGPDB

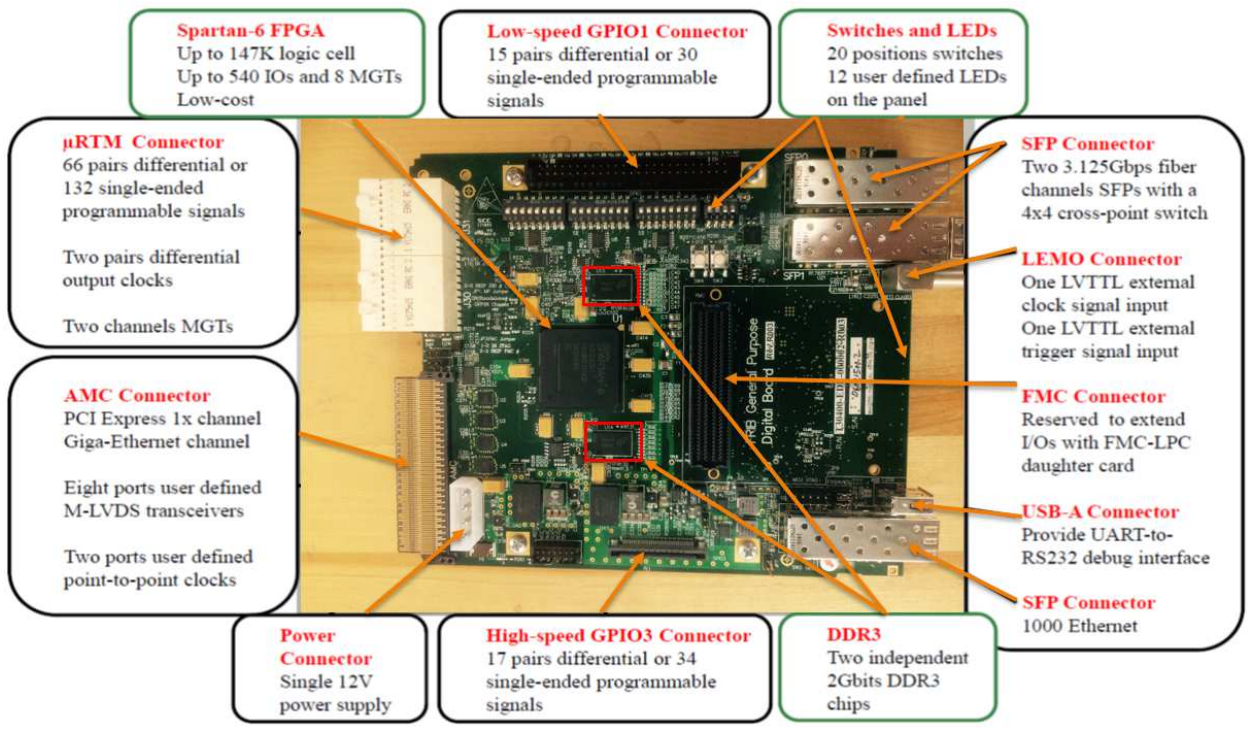

Figure 2: FGPDB peripherals and interfaces.

The FGPDB is a common digital board based on the Xilinx Spartan-6 field programmable gate array (FPGA) XC6SLX150T-FGG900 that was designed at FRIB to serve the need of different applications including LLRF, beam position monitor (BPM) and machine protection system (MPS) [3]. It is compatible with the µTCA standard and can be used either in a µTCA crate (BPM case) or in a chassis (LLRF and MPS cases). The FGPDB has two 256 MB double data rate (DDR) memory chips that are directly connected to the FPGA and controlled by the memory control block (MCB) [4]. One (Chip #1) of them is used for the Microblaze embedded system software; the other one (Chip #2) is dedicated to the waveform storage. Figure 2 shows the peripherals and interfaces of the FGPDB in more details.

## IMPLEMENTATION

The detailed implementation of the waveform storage feature will be discussed in this section.

### *Requirements*

The main purpose of the waveform storage feature is to save important internal data (e.g. the amplitude and phase information of forward/reverse/cavity signals, interlock status, etc.). The data buffer needs to hold at least one second of fast sampled data at the RF feedback control loop rate which is around 1.25 MHz (80.5 MHz / 64). The data buffer should be frozen when MPS is triggered or an internal interlock is detected. The fast data in the buffer should be able to be reorganized easily to fit the users' needs (e.g. looking at a particular variable, or at a longer time frame with lower time resolution).

The two typical use cases of the stored data are: 1. freeze the circular buffer when an interlock event happens and read out the fast data to diagnose the problem; 2. monitor one or a set of signals at a decimated rate (user settable) while the circular buffer is still running. The second case is like using an oscilloscope.

### *Memory Controller Interface*

Implementing the waveform feature involves write and read transactions to and from the memory chip. The memory interface generator (MIG) [5] provides two types of interface options when generating the MCB. One is called native port interface (NPI) or native/standard interface; the other is called advanced extensible interface (AXI). The NPI is suitable for interfacing directly to user logic. The AXI is essentially an addition of an AXI4 memory mapped slave bridge to the native interface and is suitable for interfacing with software.

For writing data into the memory, which is time critical since a new data sample (block) is generated around every 0.8 µs (1 / 1.25 MHz), the NPI interface is adopted. Figure 3 shows the protocol for loading the data into the write data FIFO (first-in-first-out) through the NPI. Figure 4 shows the protocol to entering the write request into the command FIFO. For reading data out of the memory, time is usually not a concern with data buffer frozen or lower data rate requested by the user. In this case, the AXI interface provides more flexibility in organizing the data.

MCB supports five different port configurations of up to six ports which can be read, write or bidirectional, and up to 128-bit data width (32-bit minimum). For the waveform dedicate memory (Chip #2), two 32-bit bidirectional ports are setup, one with the NPI and the other with the AXI. In practice, data is only written through the NPI and read from the AXI. However, the MIG does not directly support mixed interfaces on different ports. One has to modify the generated coded from MIG to support such configuration.

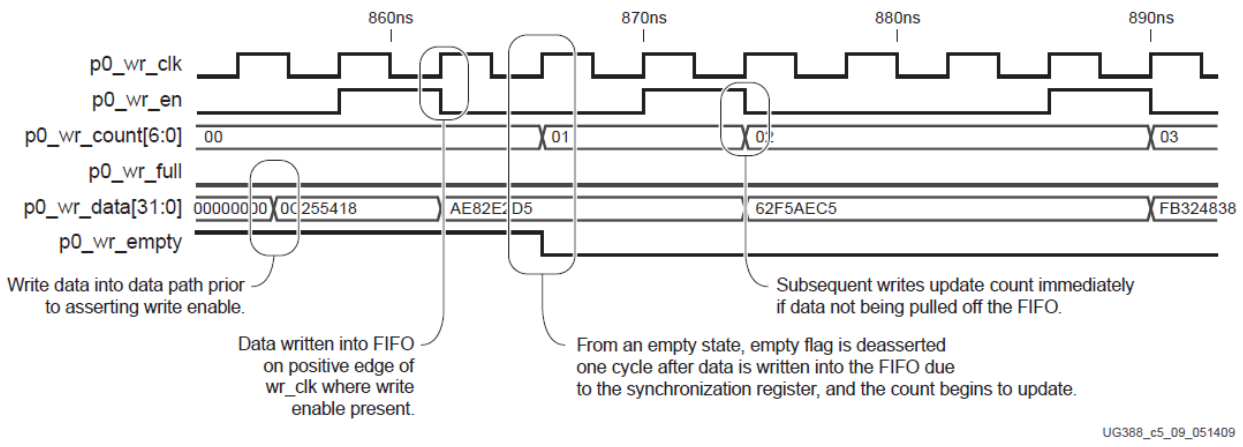

Figure 3: Loading data into write data FIFO [4].

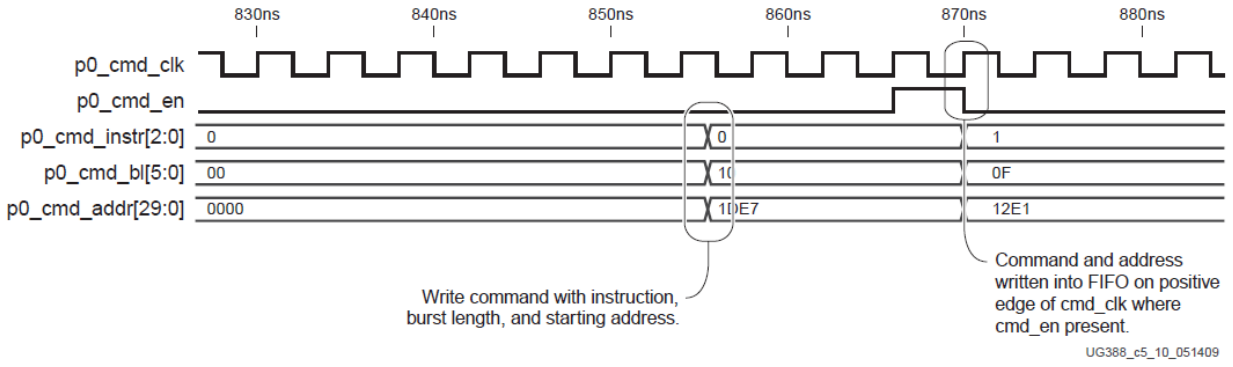

Figure 4: Entering write request into command FIFO [4].

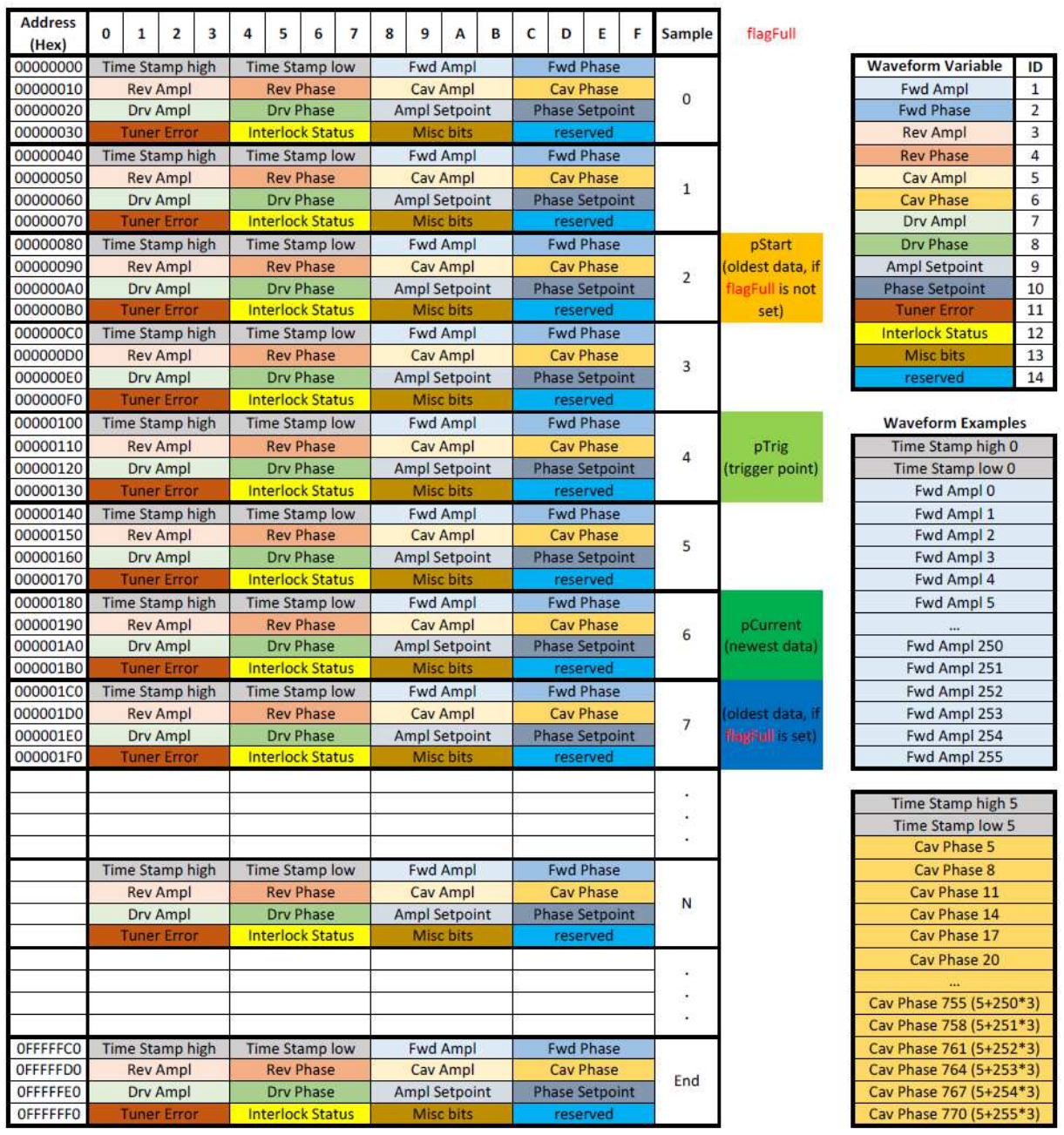

Figure 5: Data structure in the circular buffer.

### *Data Organization*

The data structure in the waveform dedicated memory is shown in Fig. 5. Currently, each sample contains sixteen 32-bit registers, which includes two time stamp registers (high and low), thirteen identified variables of interest and one reserved register. Thus the block size of a sample is 64 bytes, and the memory can hold up to 4 M samples (256 MB / 64 B) of data, which is equivalent to around 3.3 seconds ($2^{22}$ / 1.25 MHz). Even if we need to store more registers in the future, for example doubling the block size to 128 bytes, the memory is still able to hold around 1.7 seconds of data which meets the requirement of one second.

To keep track of the data in the memory, various pointers and flags are used. *pStart* is a pointer to mark the address of the starting sample of the current run; *pCurrent* is a pointer to track the address where the current sample is being written; *pTrigger* is a pointer to mark the trigger point; *flagFull* is a flag to indicate whether or not the circular buffer has been filled at least once. The *flagFull* will be set when the circular buffer is filled for the first time (*pCurrent* == *pStart*); it will be reset when the buffer starts running again for the next time. *TrigPos* is a user settable parameter that determines how many more samples to acquire after the trigger condition is met before the circular buffer freezes.

### *Data Retrieving*

For the first use case, the circular buffer is already frozen when the user tries to retrieve the data. There is no risk of data being overwritten while reading the data out. However, for the second use case the circular buffer is still running. To avoid loss of data, the fast data in the circular buffer (Chip #2) is copied over to the Microblaze memory space (Chip #1) based on user settable parameters, then the data is read out without time constraints.

This is accomplished by defining two user commands, the request waveform data (REQ_WF_DATA) command (in Fig. 6) and the read waveform (READ_WAVEFORM) command (in Fig. 7), which are based on User Datagram Protocol (UDP) communication [6].

REQ_WF_DATA Cmd Request

| Offset (Byte) | Length (Byte) | Field | Definition |
|---|---|---|---|
| 0 | 4 | Packet ID | >= 1 (incremented for each packet sent) |
| 4 | 4 | Command | 8 |
| 8 | 4 | Length | Total number of samples required. Must match Waveform NELM field |
| 12 | 4 | Decimation Factor | Number of samples to be read. DF=1 (all data read), DF=2 (one out of two) ... |

REQ_WF_DATA Cmd Response

| Offset (Byte) | Length (Byte) | Field | Definition |
|---|---|---|---|
| 0 | 4 | Packet ID | >= 1 (incremented for each packet sent) |
| 4 | 4 | Command | 8 |
| 8 | 2 | Session ID | The session ID assigned to the connection by the IOC |
| 10 | 2 | Status | Status 0 if success; !=0 if error (see LCP status codes) |
| 12 | 4 | Length | Total number of samples available to be read |
| 16 | 4 | Decimation Factor | Number of samples to be read. DF=1 (all data read), DF=2 (one out of two) ... |
| 20 | 4 | Sequence Number | Sequence Number of the package requested. Increased by 1 on each request |

Figure 6: The data format of request and response of the REQ_WF_DATA command.

When the user sends a REQ_WF_DATA command with specified *Length* ($\leq$ 8 K) and *Decimation Factor* (1 to 256), one out of every *Decimation Factor* sample(s) and a total number of *Length* samples of data are copied to the waveform buffer in Chip #2. The response packet will include a *Sequence Number* as an identifier of the newly requested waveform.

Then the user can send the READ_WAVEFORM command to read out the waveform of interest by specifying the *Waveform ID* (1 to 14). The *Offset* in the waveform buffer and the *Count* of samples to be read need to be specified as well. The *Count* is limited to 256 by the standard maximum transmission unit (MTU) of 1500. If the 9000 MTU jumbo packet is supported, this limit can extend to 2048. The payload of the response packet starts with the time stamp (two 32-bit registers) of the first data sample in the current packet followed by the *Count* number of values of the specified *Waveform ID*.

By reading multiple waveforms (with different *Waveform ID*) from the same waveform buffer (same *Sequence Number*), user can display multiple traces (up to 14) on the same time axis. By changing the *Decimation Factor*, user can easily change the time scale of the waveform. This is very similar to using an oscilloscope.

READ_WAVEFORM Cmd Request

| Offset (Byte) | Length (Byte) | Field | Definition |
|---|---|---|---|
| 0 | 4 | Packet ID | >= 1 (incremented for each packet sent) |
| 4 | 4 | Command | 3 |
| 8 | 4 | Waveform ID | ID of the waveform to be read |
| 12 | 4 | Offset | First value to be read from the waveform. Increased on each read by Count to read total length |
| 16 | 4 | Count | The number of values to return from the waveform on each read |

READ_WAVEFORM Cmd Response

| Offset (Byte) | Length (Byte) | Field | Definition |
|---|---|---|---|
| 0 | 4 | Packet ID | >= 1 (incremented for each packet sent) |
| 4 | 4 | Command | 3 |
| 8 | 2 | Session ID | The session ID assigned to the connection by the IOC |
| 10 | 2 | Status | Status 0 if success; !=0 if error (see LCP status codes) |
| 12 | 4 | Waveform ID | ID of the waveform actually read from (0xFFFFFFFF if req ID invalid) |
| 16 | 4 | Offset | The offset of the first value actually returned (0xFFFFFFFF if none) |
| 20 | 4 | Count | The number n of values returned (0 if none) |
| 24 | 4 | Sequence Number | Sequence Number of the package this data belongs to. |
| 28 | 8 | TimeStamp | Value from GTS (0 if no data returned) of first value acquired<br>[63:32]-Date/Time part of the timestamp<br>[31:0]-Sub-second part of the timestamp |
| 36 | 4*n | Waveform Values | n 32-bit values from the requested waveform |

Figure 7: The data format of request and response of the READ_WAVEFORM command.

## SUMMARY

Waveform storage feature is a useful tool for monitoring and diagnosis. The implementation meets all of the requirements and provides users with the flexibility in presenting the waveform data. The new tool will be very helpful during the next phase commissioning at FRIB. In the future, different trigger conditions could be added to the second use case providing more flexibility.